This is the author's version of an article that has been published in Proc. of ECOC 2016 – Post Deadline Paper - 42nd European Conference on Optical Communication in Düsseldorf, Deutschland. Changes were made to this version by the publisher prior to publication. The final version of record is available at https://www.vde-verlag.de/proceedings-de/454283011.html

# First Demonstration of an Automatic Multilayer Intent-Based Secure Service Creation by an Open Source SDN Orchestrator

Thomas Szyrkowiec(1), Michele Santuari(2), Mohit Chamania(1), Domenico Siracusa(2), Achim Autenrieth(1), Victor Lopez(3)

(1) ADVA Optical Networking SE, Germany tszyrkowiec@advaoptical.com
(2) CREATE-NET Research Center, Italy
(3) Telefonica I+D, Spain

**Abstract** *In this work we demonstrate an automatic intent-based encryption layer selection and configuration for a multilayer network covering IP and optical utilizing an open source SDN orchestrator. Results indicate that the processing impact of a secure channel creation is negligible.*

## Introduction

The large-scale migration of mission critical infrastructure to the public internet and data center environments has made data security an increasingly difficult proposition. Services migrating to this infrastructure also have to contend with the potential costs associated with a data breach, which are significant for sensitive applications such as banking, financial trading etc. As a result, considerable investment is made to reduce the probability of a breach, with network encryption as a key component in the overall solution. End-to-end encryption (e.g. HTTPS, SSL/TLS) can effectively address the issue of encryption in the datapath but a large number of application protocols do not natively support it and are therefore vulnerable. A typical example is the Fibre Channel protocol, which is widely used in Storage Area Networks (SANs) and does not have native encryption support[1]. With SANs now migrating beyond a dedicated physical infrastructure to a virtualized and potentially even distributed cloud infrastructure, there is a need to encrypt traffic during transmission over the network. In-flight encryption is also extensively employed by financial institutions and governmental agencies, which are extremely sensitive to data breaches.

A variety of solutions can encrypt traffic in transit and reduce the probability of a data breach. Typical techniques support encryption either at the transmission (physical) or higher layers (e.g. MACsec, IPsec) in the network stack and present trade-offs in the form of cost and performance (network latency and supported capacity). Physical layer encryption encrypts the bits entering the transmission medium and has very low latency and does not affect the throughput, while higher level encryption encrypts a frame and encapsulates it in another regular frame at the same layer, leading to reduced throughput and increased latency[2]. However, physical layer encryption requires dedicated hardware and specialized management, making it costlier (higher CAPEX/OPEX) to operate than higher level encryption, which can be performed without any specialized hardware. As a result, the choice of the encryption mechanism used, depends on the requirements of the application requesting a secure service from the network.

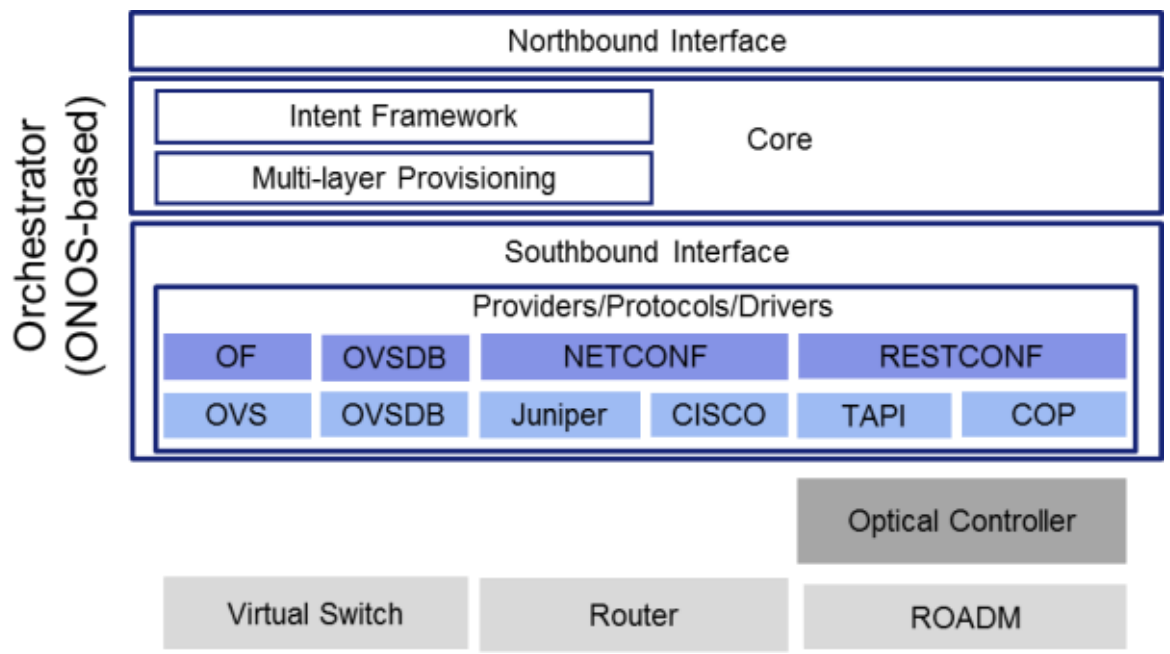

**Fig.1**: System Architecture

In this paper, we demonstrate, for the first time, the use of “application intents” to effectively move this decision complexity away from the requesting applications, making it easier for them to request secure transmission as a service. This service request is processed by an open source multilayer network orchestrator which evaluates the associated trade-offs based on the application’s requirements, and installs encryption either at the physical or at the IP layer, according to the intent expressed by the customer’s application. The orchestrator transforms the intents into a secure service, by selecting the encryption mechanism at the most appropriate layer and configuring the network devices accordingly.

## System Architecture

The intent-based multilayer orchestrator, developed in the ACINO project, is an open source effort built on top of ONOS[3] and its high level architecture is presented in Fig.1. Following the top-down approach, the intents, issued by a client, are submitted through a REST northbound interface (NBI). The orchestrator routes and compiles the intents and selects the actions that need to be taken to satisfy the intent. Those actions are translated and sent through southbound protocols to the devices that need to be configured. The devices themselves are either accessed directly or through a mediation layer like an optical controller. Proposed protocols for southbound interactions in this paper are OpenFlow, OVSDB and YANG descriptions, e.g. Control Orchestration Protocol[4] (COP) or Transport API[5] (TAPI), in combination with NETCONF or RESTCONF as a transport protocol. The optical controller is represented by ADVA's Optical Virtualization Controller (OVC). The hardware side comprises optical equipment with encryption capabilities - on a subset of the ports - as well as switches or routers which are able to install encrypted tunnels, e.g. IPsec.

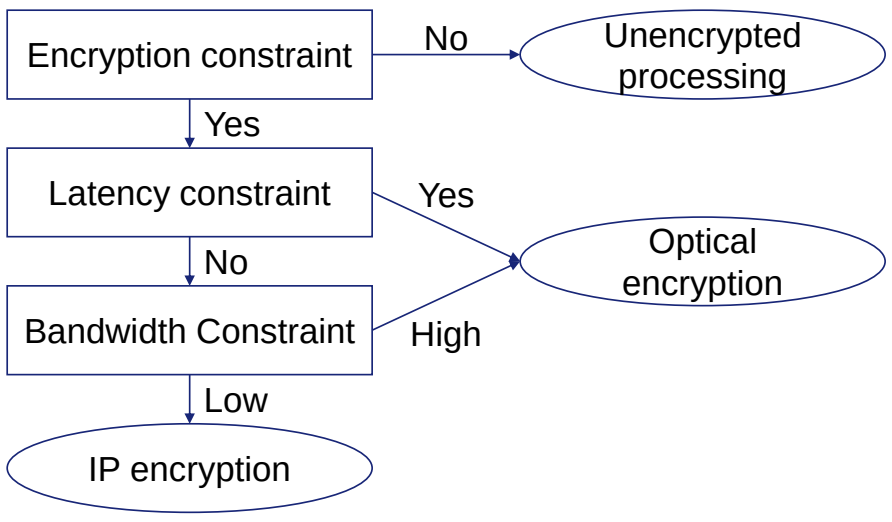

**Fig.2:** Flow chart of the decision logic

## Extensions for Encryption

The extensions of the existing orchestrator implementation[6] affected in particular the northbound interface, the intent processing and the southbound interface (SBI). In contrast to previous work, where we used the CLI, this time ONOS's REST NBI was used and extended to support the *encryption*, *latency* and *bandwidth* constraints needed for an automatic encryption assignment. The *encryption* flag indicates that the encryption processing has to take place, otherwise the intent is processed like an unencrypted request. The *latency* flag is an indicator if the traffic relies on a timely delivery. The *bandwidth* is the last criterion that impacts the choice of the best fitting encryption layer.

The ACINO intent compiler[6] was extended to handle the new constraints by applying the decision logic explained in the next section. The SBIs needed to implement new functionality to propagate the encryption information to the underlying hardware and mediation layer respectively. In the case of switches, a GRE tunnel setup routine was added to establish an IP layer encryption leveraging the OVSDB protocol.

The COP[4] protocol (and the associated driver) was augmented by introducing a flag to indicate encryption to the underlying (optical) controller:

*augment "/cop:calls/cop:call" {*
*container encryption { presence "encrypt call"}}*

Finally, the optical controller, i.e. OVC, itself needed to handle the encryption requests, to configure the ports and to setup the lightpath.

## Decision Logic

As mentioned earlier, the parameters that are evaluated, when choosing an encryption mechanism, include the encryption flag, the latency flag and a bandwidth value. As shown in Fig.2, the encryption flag is checked first to verify that the traffic is privacy sensitive. If this is the case, the latency constraint is evaluated next, else the processing for unencrypted traffic is applied. Since the IP layer encryption introduces more delay[2] than a physical encryption, the lower layer is chosen for latency sensitive traffic. Finally, the bandwidth is checked in order to determine if an IP-based solution can satisfy the demand or if physical layer encryption, working at line speed[7], is required. Depending on this parameter the final decision is made and the required actions are initiated.

## Experimental Validation

The presented system architecture was implemented and evaluated with commercial hardware in a lab testbed (Fig.3). We used three PCs to host the software components. Two machines hosted an Open vSwitch (OVS) instance each, including a virtual host interface for end to end verification. The other computer ran the controllers, i.e. ONOS and OVC, and was connected to the optical equipment as well as the OVS instances through a management network.

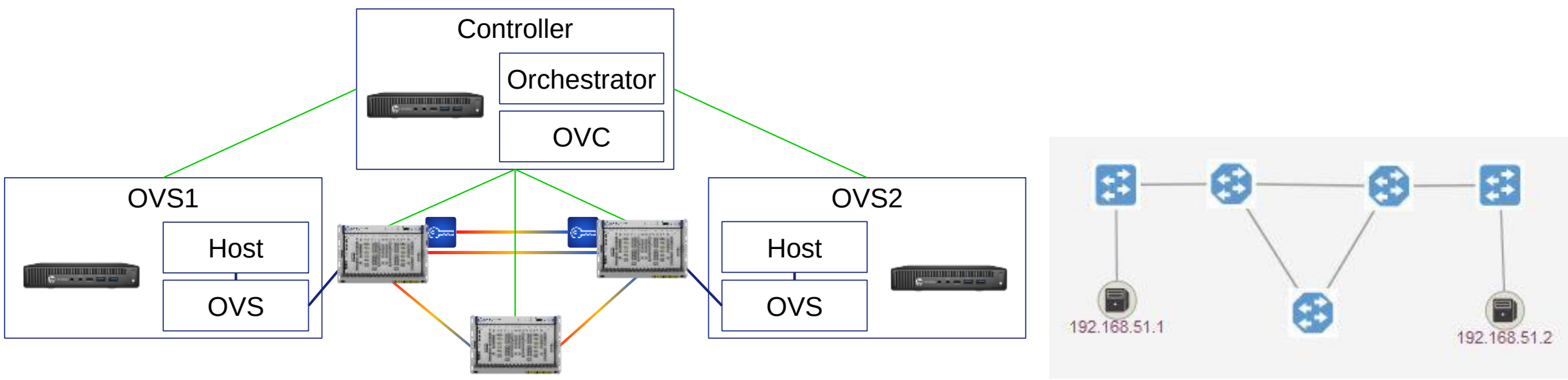

**Fig.3:** Testbed setup and ONOS's view of the testbed

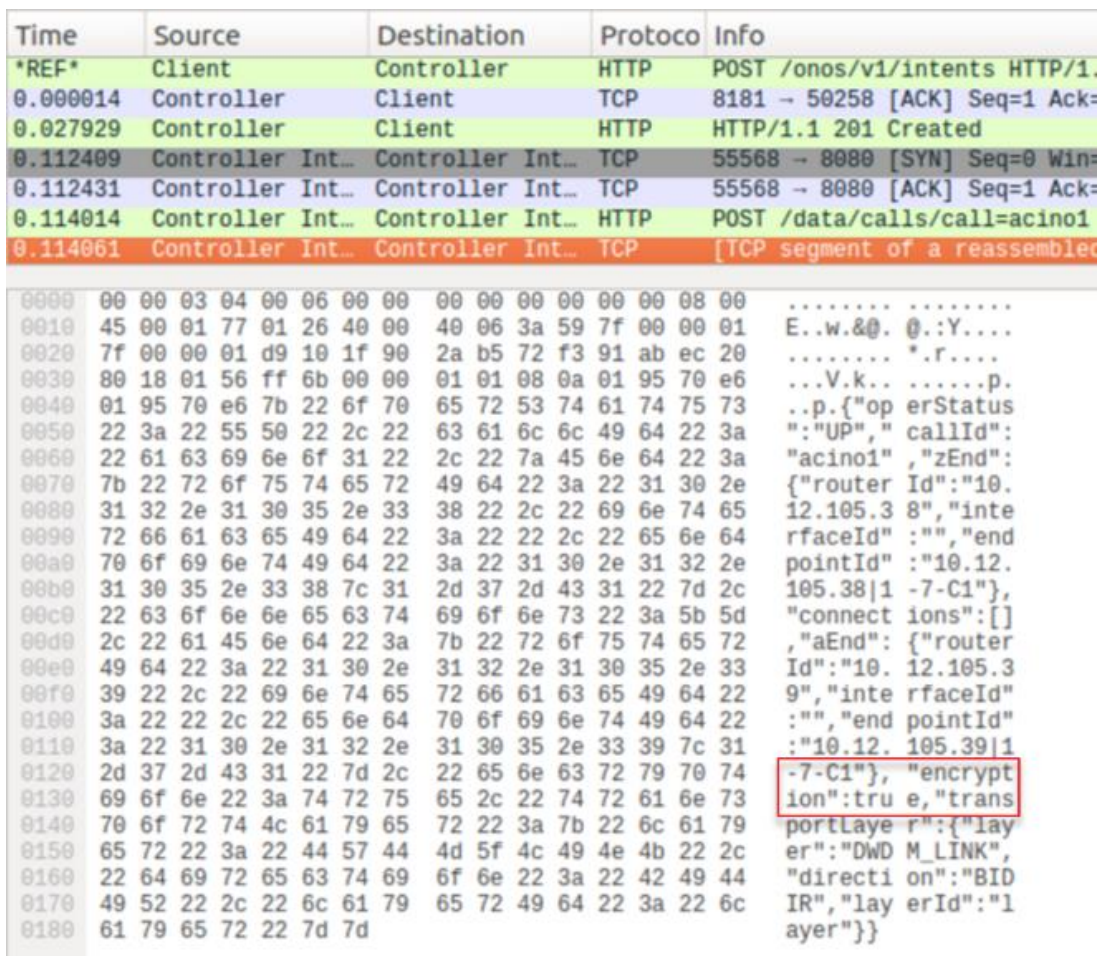

**Fig.4:** Physical encryption setup trace

The testbed included an ADVA FSP3000 ROADM ring consisting of three nodes. Two of them were equipped with AES cards[7] which encrypt all traffic on the physical layer. The secret key for both encryption cards was configured by the administrator. For the IP encryption ONOS triggered the creation of a GRE tunnel between both OVS instances. The tunnel became active after the lightpath was setup between the network ports on both sides. We evaluated two scenarios that requested secure service. The best suited layer for the encryption was chosen automatically by the orchestrator applying the previously described decision logic.

## Measurements

We conducted experimental tests for both layers of encryption, i.e. physical as well as IP.

In the first scenario we used an intent that requires an encryption, is latency sensitive and the requested bandwidth was 1 Mbit/s between the host on OVS1 and the host on OVS2. One example for such traffic could be a real-time audio conference. Based on those constraints the ACINO orchestrator decides to use optical encryption because of the latency. The detailed Wireshark trace is shown in Fig.4. First the client sends a request to the orchestrator located at the controller node. The intent is confirmed and the processing is started. Next an internal controller message is sent from ONOS to the OVC (port 8080) requesting an optical lightpath between the client ports of the ROADMs (10.12.105.38 & 39). The content of the request is shown in the lower part of Fig.4. All necessary fields for a service setup are included. The encryption flag that was added as part of this work is highlighted in the picture. The processing time, starting with the intent request at the NBI and ending with the setup message for an encrypted optical tunnel at the SBI, takes less than 120 ms.

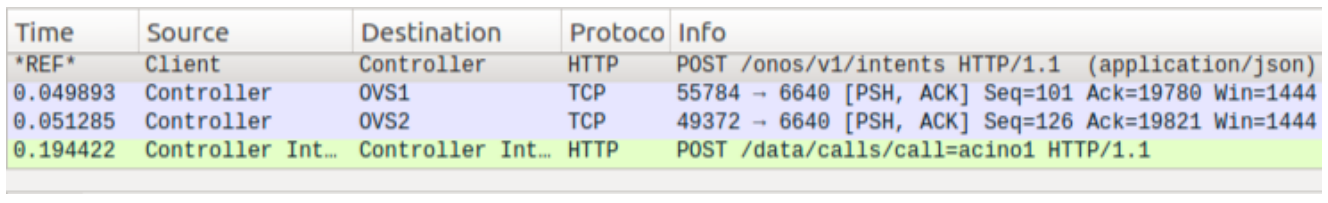

**Fig.5:** IP encryption setup trace

In the second scenario we used a request for encryption which only required a small bandwidth of 1 Mbit/s without being latency sensitive. According to our decision algorithm this leads to an IP layer encryption. The endpoints remained the same as before. The Wireshark trace of this run is depicted in Fig.5. We boiled the trace down to the most important packets, starting again with the initial intent. After that we see a difference in the course of messages. The controller sends OVSDB configuration messages to OVS1 and OVS2 to initialize the GRE tunnel on both sides. After that the tunnel setup in between is triggered. The COP request, not presented here due to space constraints, is missing the encryption flag according to the "presence" definition in YANG. As soon as the optical lightpath is created the encrypted IP tunnel can be used to transmit privacy sensitive data. As we can see here the processing time has slightly grown to about 200 ms due to the additional steps for the configuration of both virtual switches as well as the related messaging.

## Conclusions

In this work we have demonstrated for the first time an automatic intent-based encryption selection and configuration for a multilayer network, covering IP and optical, by an orchestrator. The results indicate that the processing time is negligible to the lightpath setup, which takes a few seconds per hop. In the future we expect to automate the key setup needed for the encryption.

## Acknowledgements

This research has received funding from the European Commission within the H2020 Programme, ACINO project, Grant Number 645127.